\documentclass[]{spie}  

\usepackage{amsmath,amsfonts,amssymb}
\usepackage{graphicx}
\usepackage{subcaption}
\usepackage[colorlinks=true, allcolors=blue]{hyperref}
\usepackage{siunitx}

\title{The Design and Performance Characteristics of the NRL4 ASIC Developed for the COSI Small Explorer Gamma-ray Satellite}

\author[a]{Jarred M. Roberts}
\author[b]{Clio Sleator}
\author[b]{Eric Wulf}
\author[a]{Steven E. Boggs}
\author[d]{Gianluigi De Geronimo}
\author[c]{Alex Lowell}
\author[c]{Brent Mochizuki}
\author[c]{John A. Tomsick}

\affil[a]{University of California San Diego, 9500 Gilman Dr. La Jolla, CA 92093, USA}
\affil[b]{U.S. Naval Research Laboratory, 4555 Overlook Ave. S.W. Washington, DC 20375, USA}
\affil[c]{Space Sciences Laboratory, 7 Gauss Way, Berkeley, CA 94720, USA}
\affil[d]{DG Circuits, Syosset, NY 11791, USA}

\authorinfo{Further author information: (Send correspondence to J.M.R)\\J.M.R.: E-mail: jmroberts@physics.ucsd.edu}

\begin{document}
\maketitle

\begin{abstract}
Next-generation gamma-ray observatories aim to enable precision measurements in high-energy astrophysics using advanced semiconductor detector technologies. Meeting the scientific requirements of modern instruments demands detector systems that provide high spatial and spectral resolution across large detection areas, with strict limits on power consumption and mass. These needs drive innovation in front-end electronics and mixed-signal processing to support compact detector electrode geometries.

Application-specific integrated circuits (ASICs) are essential in front-end readout electronics, enabling high-channel-density and low-power systems, while maintaining low-noise performance suitable for space-based instruments and balloon-borne payloads. The NRL4 (Naval Research Laboratory 4) is a recently developed 32-channel front-end ASIC featuring low-power, low-noise channels consisting of charge-sensitive preamplifiers, 4 configurable gain settings, dual configurable shapers for optimized timing and energy resolution, trimmable per-channel discrimination, time-to-amplitude conversion (TAC), and peak-detect output.

The NRL4 has been integrated with a high-purity germanium (HPGe) dual-sided strip detector with a \SI{1.16}{\milli\meter} strip pitch. Energy resolution of \SI{3}{\kilo\electronvolt} full width at half maximum (FWHM) at \SI{59.54}{\kilo\electronvolt} was achieved with a gain of \SI{18.4}{\milli\volt\per\femto\coulomb} and a slow shaper peaking time of \SI{2}{\micro\second}. Preliminary results from ongoing research demonstrate the suitability of the NRL4 for high-resolution, low-power gamma-ray spectroscopy for ground and space-based missions. 
\end{abstract}



\begin{figure}[t]
\centering
\begin{subfigure}[t]{0.68\linewidth}
    \centering
    \includegraphics[width=\linewidth]{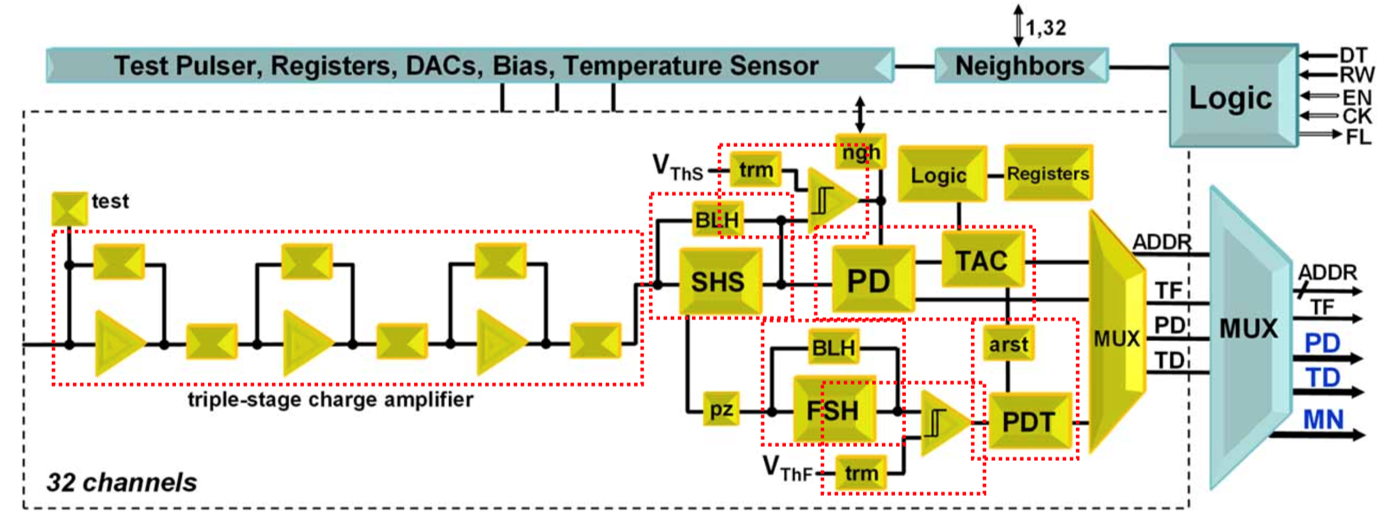}
\end{subfigure}
\hfill
\begin{subfigure}[t]{0.3\linewidth}
    \centering
    \includegraphics[width=\linewidth]{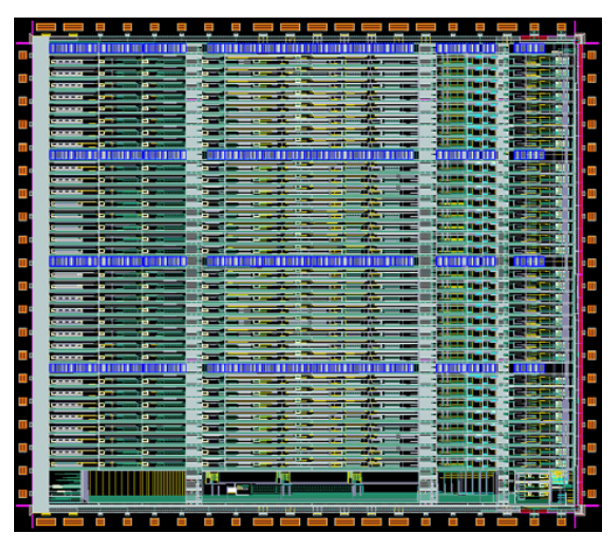}
\end{subfigure}
\vspace{0.8\baselineskip}
\caption{(\textit{Left}) Simplified block diagram of the major functions of the NRL4 ASIC. Each of the 32 channels is amplified by a charge-sensitive preamplifier, and the signal is processed using two parallel shapers: a second-order fast shaper and a fifth-order slow shaper, both with baseline stabilization and trimmable thresholds. (\textit{Right}) Physical layout of the NRL4 ASIC.}
\label{fig1}
\end{figure}

\begin{figure}[t]
\centering
\begin{subfigure}[t]{0.49\linewidth}
    \centering
    \includegraphics[width=\linewidth]{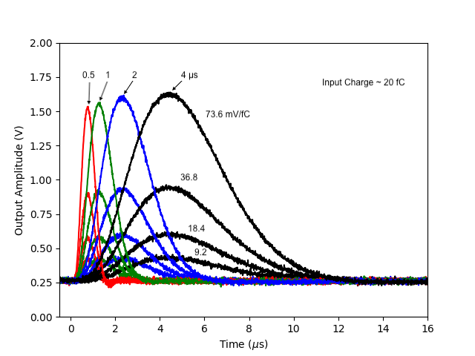}
\end{subfigure}
\hfill
\begin{subfigure}[t]{0.49\linewidth}
    \centering
    \includegraphics[width=\linewidth]{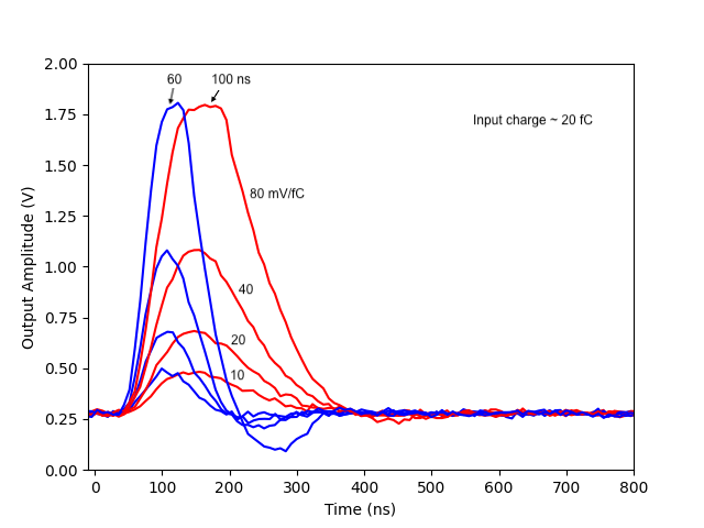}
\end{subfigure}
\caption{(\textit{Left}) Gain and peaking time settings of the NRL4 slow-shaping circuit for precision energy measurements, shown for an input charge of \SI{20}{\femto\coulomb}. (\textit{Right}) Gain and peaking time settings of the NRL4 fast-shaping circuit for precision timing measurements, also for an input charge of \SI{20}{\femto\coulomb}.}
\label{fig2}
\end{figure}

\section{Introduction}
\label{sec:introduction}

The excellent energy resolution of HPGe makes it an ideal semiconductor detector medium for gamma-ray physics and astrophysics \cite{Tomsick_ICRC_2023, Siegert2020, Lowell2017, Shih2012, Roberts_2025, Greiner2012, Boggs2005}, industry \cite{Phlips2002}, homeland security and nuclear non-proliferation \cite{wulf2003}, and nuclear physics \cite{Rumaiz2014} applications aiming to optimize the spectroscopic capabilities of their instruments. In astrophysics applications, the pitch of the aluminum strip electrodes used to collect the charge carriers is directly correlated with the angular resolution capabilities of the instrument and, therefore, the science capabilities of the gamma-ray telescope; HPGe strip detectors can be utilized in both Compton telescopes and collimation systems. As the science motivates increasingly fine-pitched strip detectors, the resulting increased channel density produces significant challenges to space and balloon-borne instruments that are constrained by strict mass, volume, and power budgets. A unique requirement in developing readout electronics for HPGe detectors is the ability to accurately measure the time of the charge collection at the anode and cathode strips; HPGe detectors are designed to fully absorb photons, are on the order of a few \si{\centi\meter} thick, and therefore have charge drift times on the order of \SI{100}{\nano\second}. The development of ASICs capable of low-power, high channel density readout, with precision energy and timing reconstruction, is strongly motivated in this field.

COSI is a wide-field-of-view Compton gamma-ray imager, spectrometer, and polarimeter that employs an array of 16 HPGe dual-sided strip detectors arranged in a $2\times2\times4$ configuration. These detectors require low-noise, low-power, low-footprint front-end readout electronics, with precision energy and timing capabilities. The HPGe detectors have 64 strips per detector side and a thickness of approximately \SI{150}{\milli\meter}. COSI is a NASA-funded Small Explorer-class mission, slated for launch into a low-Earth near-equatorial orbit in 2027. The NRL4 ASIC design, ideal for such detectors, serves as the flight ASIC for the COSI instrument, ensuring high-performance detection and data processing for gamma-ray observations.


\section{Instrumentation Overview}

\begin{figure}[t]
\centering
\begin{subfigure}[t]{0.48\linewidth}
    \centering
    \includegraphics[width=\linewidth]{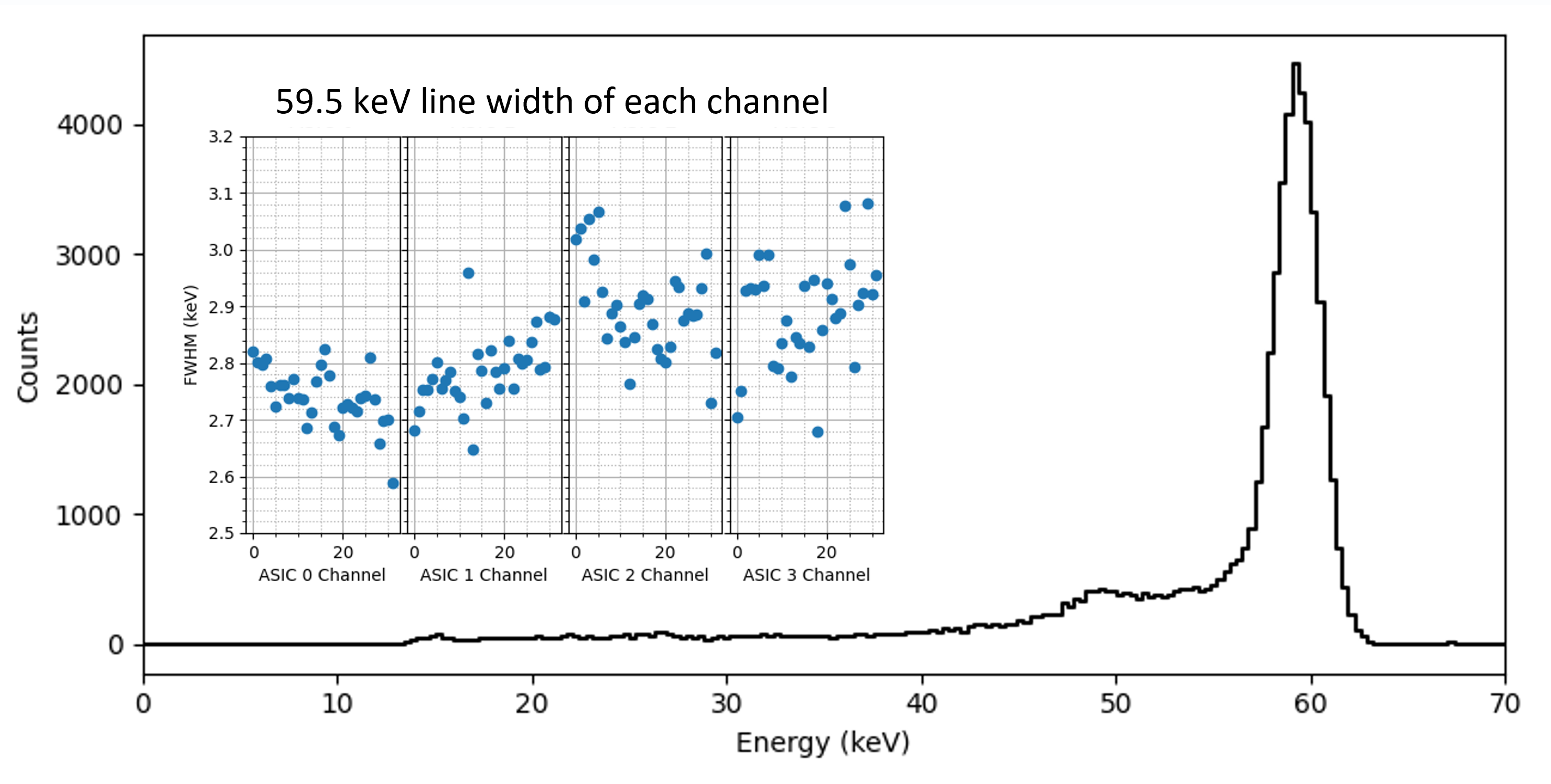}
\end{subfigure}
\hfill
\begin{subfigure}[t]{0.48\linewidth}
    \centering
    \includegraphics[width=\linewidth]{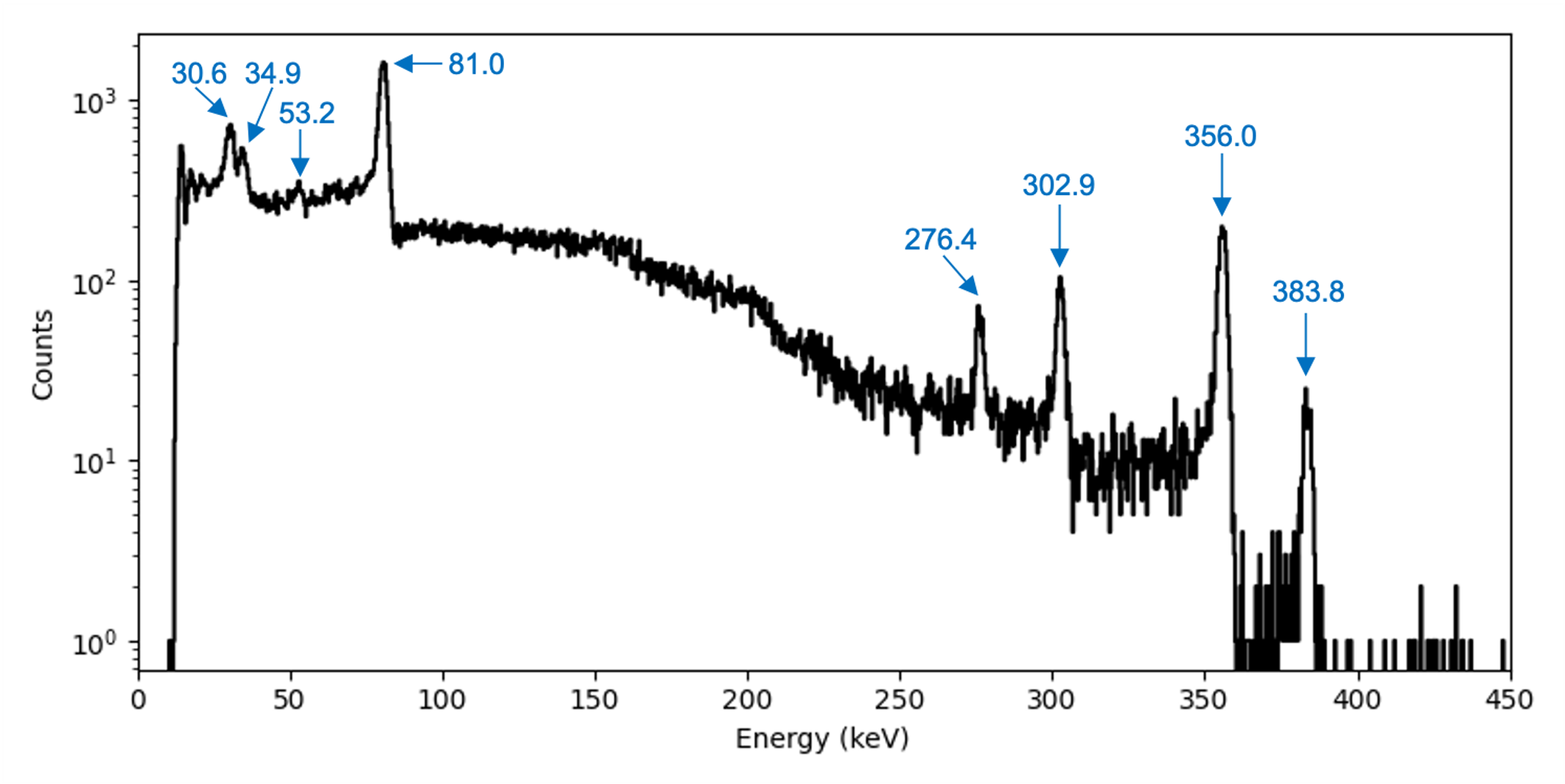}
\end{subfigure}
\caption{(\textit{Left}) Measured energy resolution of 64 low-voltage detector strips and 64 high-voltage detector strips, illuminated by a $^{241}\mathrm{Am}$ source. (\textit{Right}) $^{133}\mathrm{Ba}$ spectrum measured by a single low-voltage NRL4 channel, with labeled photopeaks demonstrating the ability to resolve the 30.6 and \SI{34.9}{\kilo\electronvolt} photopeaks.}
\label{fig3}
\end{figure}

\begin{figure}[t]
\centering
\begin{subfigure}[t]{0.45\linewidth}
    \centering
    \includegraphics[width=\linewidth]{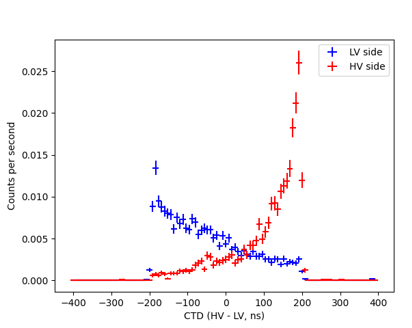}
\end{subfigure}
\hfill
\begin{subfigure}[t]{0.54\linewidth}
    \centering
    \includegraphics[width=\linewidth]{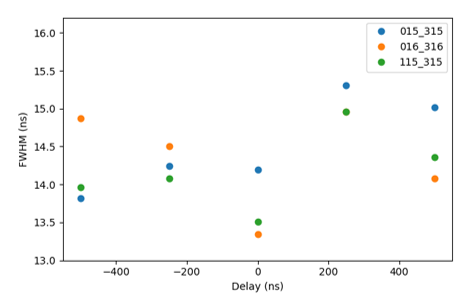}
\end{subfigure}
\caption{(\textit{Left}) Collection time difference (CTD) of the high-voltage and low-voltage sides of an HPGe detector illuminated by a $^{57}\mathrm{Co}$ source. (\textit{Right}) Measured coincidence timing resolution of the NRL4 ASIC between pixel pairs (shown in the legend) on opposite sides of an HPGe detector, using the internal pulser of the NRL4.}
\label{fig4}
\end{figure}

The NRL4 is the fourth-generation ASIC in the Naval Research Laboratory (NRL) family, based on the architecture of the NRL1 \cite{Wulf2020} and NCIASIC2 \cite{wulf2003}. Similar to the NRL1, the NRL4 is a 32-channel front-end ASIC with charge-sensitive preamplifiers configurable for both positive and negative signal polarities, and with four gain settings: 9.2, 18.4, 36.8, and \SI{73.6}{\milli\volt\per\femto\coulomb}. The total power draw of the ASIC is \SI{218}{\milli\watt}, or approximately \SI{6.8}{\milli\watt} per channel. A simplified block diagram of the NRL4 and an image of the ASIC layout are shown in Fig.~\ref{fig1}. The ASIC includes an analog monitor that can be programmed to output any of the 32 channels after either the fast or slow shaper, which is useful for debugging and signal-to-noise analysis.

The NRL4 routes the preamplified signal to two independent shaper circuits: a fast, second-order shaper with configurable peaking times of 60 and \SI{100}{\nano\second}, optimized for timing reconstruction, and a slow, fifth-order shaper optimized for energy reconstruction with configurable peaking times of 0.5, 1, 2, and \SI{4}{\micro\second} (Fig. \ref{fig2}). Each shaping circuit incorporates baseline stabilization for improved noise suppression and trimmable thresholds.

Each channel includes a time-to-amplitude converter (TAC) for precision timing measurement and a peak detector with analog memory for slow shaper pulse amplitude measurement. Interaction depth is determined from the collection time difference (CTD) between strips on opposite sides of the detector \cite{rogers2025}. The NRL4 also supports nearest-neighbor readout: it can trigger sub-threshold readout of adjacent channels, and if channels 1 or 32 are triggered, it can signal an adjacent ASIC to process the neighboring strip. This improves event reconstruction when total charge from an event is shared between multiple strips.

To characterize NRL4 performance, HPGe strip detectors have been tested and instrumented with readout electronics at the Space Sciences Laboratory (SSL) and at the U.S. Naval Research Laboratory \cite{Sleator_2023}. These cryogenically cooled (to approximately \SI{80}{\kelvin}) detectors have a \SI{1.16}{\milli\meter} strip pitch, 64 strips per side, and 4,096 pixel cross-sections. The detectors' guard ring channel is read out using a dedicated NRL4 ASIC on each side of the detector, requiring six ASICs to read out all strips of a single GeD. These HPGe detectors have been developed specifically for COSI.


\section{Experimental Results}

\begin{figure}[t]
\centering
\begin{subfigure}[t]{0.43\linewidth}
    \centering
    \includegraphics[width=\linewidth]{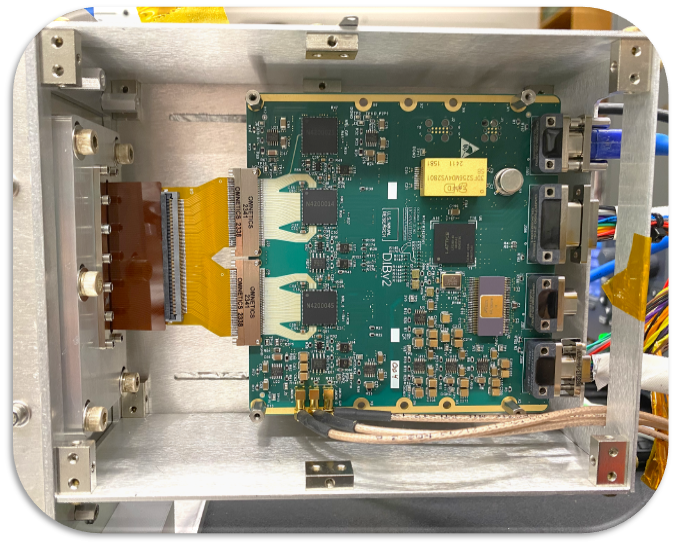}
\end{subfigure}
\hfill
\begin{subfigure}[t]{0.54\linewidth}
    \centering
    \includegraphics[width=\linewidth]{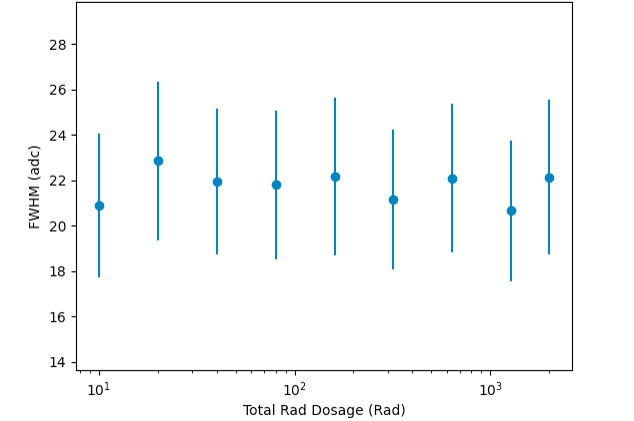}
\end{subfigure}
\vspace{1.0\baselineskip}
\caption{(\textit{Left}) Detector Interface Board Version 2 (DIBv2) instrumented with an HPGe strip detector. (\textit{Right}) Energy resolution of the NRL4 ASICs as a function of radiation dose.}
\label{fig5}
\end{figure}

Multiple HPGe strip detectors were illuminated with various radioactive sources while instrumented with the NRL4 readout electronics. As shown in Fig.~\ref{fig3}, the energy resolution was measured for each channel at \SI{59.54}{\kilo\electronvolt} using an $^{241}\mathrm{Am}$ source, a gain setting of \SI{18.4}{\milli\volt\per\femto\coulomb} and a shaping time of \SI{2}{\micro\second}. Most channels achieved better than \SI{3}{\kilo\electronvolt} FWHM, with several achieving better than \SI{2.7}{\kilo\electronvolt} FWHM. Edge channels 0--5 and 26--31, which historically exhibited poorer energy resolution in earlier ASIC generations, show improved performance in the NRL4 due to reduced resistance in the edge channel traces in the ASIC layout.

Timing measurements are ongoing to quantify the resolution of the new circuit based on a low--time-walk peak-timing method on a dedicated fast shaper \cite{Wulf2020}. Preliminary NRL4 data, obtained with the ASIC connected to an HPGe detector load and using the internal pulser to generate time-delayed pulses on strips on opposite detector sides (Fig.~\ref{fig4}), demonstrate coincidence timing resolutions of $14.5 \pm 1\,\si{\nano\second}$. Figure~\ref{fig4} also shows the CTD distribution of an HPGe detector illuminated by a $^{57}\mathrm{Co}$ source.

The NRL4 ASIC has been characterized using flight-like, custom-developed readout electronics with HPGe detectors, meeting performance requirements for the COSI space mission. It has also demonstrated suitability for other semiconductor detector technologies, including CZT detectors for the XL-Calibur X-ray telescope \cite{XLCalibur}, silicon detectors for the APT gamma-ray observatory \cite{Buckley2024}, and SiPMs coupled to BGO scintillators for the anti-coincidence shield system onboard the COSI satellite payload.


\section{Conclusion}
A next-generation NRL ASIC has been developed, fabricated, and tested using custom readout electronics with an HPGe detector featuring a \SI{1.16}{\milli\meter} strip pitch. The NRL4 demonstrated energy-resolution performance of \SI{3}{\kilo\electronvolt} FWHM at \SI{59.54}{\kilo\electronvolt} with a gain setting of \SI{18.4}{\milli\volt\per\femto\coulomb} and a shaping time of \SI{2}{\micro\second}. Preliminary timing measurements indicate a resolution of $14.5 \pm 1\,\si{\nano\second}$. Radiation-hardened, front-end ASIC-interfacing readout electronics have also been developed and have maintained performance under radiation doses consistent with those expected during a two-year mission in low-Earth orbit.


\section*{Acknowledgments}

This work was sponsored by NASA (United States) NNH18ZDA001N APRA and NASA Explorer's contract 80GSFC21C0059 and grant number 80NSSC19K1389.


\bibliography{report} 

\begin{thebibliography}{10}

\bibitem{Tomsick_ICRC_2023}
Tomsick, J., Boggs, S., Zoglauer, A., Hartmann, D.~H., Ajello, M., Burns, E., Fryer, C., Karwin, C., Kierans, C., Lowell, A., Malzac, J., Roberts, J., Saint-Hilaire, P., Shih, A., Siegert, T., Sleator, C., Takahashi, T., Tavecchio, F., Wulf, E., Beechert, J., Gulick, H., Joens, A., Lazar, H., Neights, E., Martinez~Oliveros, J.~C., Matsumoto, S., Melia, T., Yoneda, H., Amman, M., Bal, D., von Ballmoos, P., Bates, H., Böttcher, M., Bulgarelli, A., Cavazzuti, E., Chang, H.-K., Chen, C., Chu, C.-Y., Ciabattoni, A., Costamante, L., Dreyer, L., Fioretti, V., Fenu, F., Gallego, S., Ghirlanda, G., Grove, E., Huang, C.-Y., Jean, P., Khatiya, N., Knödlseder, J., Kraus, M., Leising, M., Lewis, T., Lommler, J., Marcotulli, L., Martinez~Castellanos, I., Mittal, S., Negro, M., Al~Nussirat, S., Nakazawa, K., Oberlack, U., Palmore, D., Panebianco, G., Parmiggiani, N., Pike, S., Rogers, F., Schutte, H., Sheng, Y., Smale, A., Smith, J.~R., Trigg, A., Venters, T., Watanabe, Y., and Zhang, H., ``{The Compton Spectrometer and
  Imager},'' {\em PoS}~{\bf ICRC2023},  745 (2023).

\bibitem{Siegert2020}
Siegert, T. et~al., ``Imaging the 511 kev positron annihilation sky with cosi,'' {\em Astrophysical Journal}~{\bf 45},  897 (2020).

\bibitem{Lowell2017}
Lowell, A.~W. et~al., ``Polarimetric analysis of the long duration gamma-ray burst grb 160530a with the balloon borne compton spectrometer and imager,'' {\em Astrophysical Journal}~{\bf 119},  848 (2017).

\bibitem{Shih2012}
Shih, A.~Y. et~al., ``The gamma-ray imager/polarimeter for solar flares (grips),'' in [{\em Space Telescopes and Instrumentation 2012: Ultraviolet to Gamma Ray}{\nolinebreak\hspace{0.1em}]},  (2012).

\bibitem{Roberts_2025}
Roberts, J.~M., Boggs, S., Siegert, T., Tomsick, J.~A., Ajello, M., von Ballmoos, P., Beechert, J., Cangemi, F., Gallego, S., Jean, P., Karwin, C., Kierans, C., Lazar, H., Lowell, A., Martinez~Castellanos, I., Pike, S., Sleator, C., Sheng, Y., Yoneda, H., and Zoglauer, A., ``Imaging and spectral fitting of bright gamma-ray sources with the cosi balloon payload,'' {\em The Astrophysical Journal}~{\bf 979},  116 (jan 2025).

\bibitem{Greiner2012}
Greiner, J., Mannheim, K., Aharonian, F., Ajello, M., Balasz, L.~G., Barbiellini, G., Bellazzini, R., Bishop, S., Bisnovatij-Kogan, G.~S., Boggs, S., Bykov, A., DiCocco, G., Diehl, R., Elsässer, D., Foley, S., Fransson, C., Gehrels, N., Hanlon, L., Hartmann, D., Hermsen, W., Hillebrandt, W., Hudec, R., Iyudin, A., Jose, J., Kadler, M., Kanbach, G., Klamra, W., Kiener, J., Klose, S., Kreykenbohm, I., Kuiper, L.~M., Kylafis, N., Labanti, C., Langanke, K., Langer, N., Larsson, S., Leibundgut, B., Laux, U., Longo, F., Maeda, K., Marcinkowski, R., Marisaldi, M., McBreen, B., McBreen, S., Meszaros, A., Nomoto, K., Pearce, M., Peer, A., Pian, E., Prantzos, N., Raffelt, G., Reimer, O., Rhode, W., Ryde, F., Schmidt, C., Silk, J., Shustov, B.~M., Strong, A., Tanvir, N., Thielemann, F.-K., Tibolla, O., Tierney, D., Trümper, J., Varshalovich, D.~A., Wilms, J., Wrochna, G., Zdziarski, A., and Zoglauer, A., ``{GRIPS - Gamma-Ray Imaging, Polarimetry and Spectroscopy},'' {\em Experimental Astronomy}~{\bf 34}(2),  551--582
  (2012).

\bibitem{Boggs2005}
Boggs, S., Bandstra, M., Bowen, J., Coburn, W., Lin, R., Wunderer, C., Zoglauer, A., Amman, M., Luke, P., Jean, P., and von Ballmoos, P., ``Performance of the nuclear compton telescope,'' {\em Experimental Astronomy}~{\bf 20}(1),  387--394 (2005).

\bibitem{Phlips2002}
Phlips, B.~F. et~al., ``Development of germanium strip detectors for environmental remediation,'' {\em IEEE Transactions on Nuclear Science}~{\bf 49},  597--600 (2002).

\bibitem{wulf2003}
Wulf, E.~A. et~al., ``Germanium strip detector compton telescope using three-dimensional readout,'' {\em IEEE Transactions on Nuclear Science}~{\bf 50},  1182--1189 (2003).

\bibitem{Rumaiz2014}
Rumaiz, A.~K. et~al., ``A monolithic segmented germanium detector with highly integrated readout,'' {\em IEEE Transactions on Nuclear Science}~{\bf 61}(6),  3721--3726 (2014).

\bibitem{Wulf2020}
Wulf, E.~A., Hou, W., Geronimo, G.~D., Roberts, J.~M., Boggs, S.~E., and Phlips, B.~F., ``Front-end asic for germanium strip detectors,'' {\em Nuclear Instruments and Methods in Physics Research Section A}~{\bf 954},  161230 (2020).

\bibitem{rogers2025}
Rogers, F., Pike, S., Nussirat, S.~A., Lowell, A., Boggs, S., Anthony-Petersen, R., Broadbent, E., Gerber, J., Haight, S.~E., Kierans, C., Mochizuki, B., Patel, P., Roberts, J., Shih, A.~Y., Sleator, C., Szornel, J., Tomsick, J.~A., Valluvan, A., Wieber, M., and Zoglauer, A., ``Calibration of germanium double-sided strip detectors for the compton spectrometer and imager (cosi) satellite,'' Manuscript in preparation for submission to SPIE (2025).

\bibitem{Sleator_2023}
Sleator, C.~C., Wulf, E.~A., Lowell, A., Mochizuki, B., Joens, A., de~Geronimo, G., Roberts, J.~M., Smith, J., Davis, J.~M., Johnson-Rambert, M., and Tomsick, J.~A., ``Front-end asic spectral and timing performance on a high purity germanium strip detector,'' in [{\em 2023 IEEE Nuclear Science Symposium, Medical Imaging Conference and International Symposium on Room-Temperature Semiconductor Detectors (NSS MIC RTSD)}{\nolinebreak\hspace{0.1em}]},   1--2 (2023).

\bibitem{XLCalibur}
Iyer, N., Kiss, M., Pearce, M., Stana, T.-A., Awaki, H., Bose, R., Dasgupta, A., {De Geronimo}, G., Gau, E., Hakamata, T., Ishida, M., Ishiwata, K., Kamogawa, W., Kislat, F., Kitaguchi, T., Krawczynski, H., Lisalda, L., Maeda, Y., Matsumoto, H., Miyamoto, A., Miyazawa, T., Mizuno, T., Rauch, B., Cavero, N.~R., Sakamoto, N., Sato, J., Spooner, S., Takahashi, H., Takeo, M., Tamagawa, T., Uchida, Y., West, A., Wimalasena, K., and Yoshimoto, M., ``The design and performance of the xl-calibur anticoincidence shield,'' {\em Nuclear Instruments and Methods in Physics Research Section A: Accelerators, Spectrometers, Detectors and Associated Equipment}~{\bf 1048},  167975 (2023).

\bibitem{Buckley2024}
Buckley, J.~H., Buhler, J., and Chamberlain, R.~D., ``{The Advanced Particle-astrophysics Telescope (APT): Computation in Space},'' in [{\em {21th ACM International Conference on Computing Frontiers}}{\nolinebreak\hspace{0.1em}]},  (2024).

\end{thebibliography}
\bibliographystyle{spiebib} 
\end{document}